\documentstyle[prl,aps,multicol]{revtex}

\draft

\author{Jennifer~L.~Dodd, Michael~A.~Nielsen, Michael~J.~Bremner, 
and Robert~T.~Thew}
\title{Universal quantum computation and simulation using any
entangling Hamiltonian and local unitaries}
\address{Centre for Quantum Computer Technology, Department of
Physics, University of Queensland 4072, Australia} \date{\today}

\begin{document}

\bibliographystyle{prsty}

\pagestyle{plain}
\pagenumbering{arabic}

\maketitle

\begin{abstract}

What interactions are sufficient to simulate {\em arbitrary} quantum
dynamics in a composite quantum system?  We provide an {\em efficient}
algorithm to simulate any desired two-body Hamiltonian evolution using
any fixed two-body entangling $n$-qubit Hamiltonian and local
unitaries.  It follows that universal quantum computation can be
performed using {\em any} entangling interaction and local unitary
operations.

\end{abstract}


\begin{multicols}{2}[]
\narrowtext
%
%

A central goal of quantum physics is to understand and control quantum
dynamics.  Recently, the emergence of fields such as quantum
control\cite{Rabitz00a}, laser cooling\cite{Wieman99a}, quantum
communication and quantum computation\cite{Nielsen00a,Preskill98c} has
focused efforts to understand and control quantum dynamics at the
single quantum level.

Our interest is in the dynamics of composite quantum systems.  An
especially important example of such a system is a quantum computer,
which is a composite of a large number of two-level quantum systems
(qubits).  We wish to determine which interactions are sufficient for
the simulation of {\em arbitrary} quantum dynamics in such a system.
Our results demonstrate equivalence between this property of {\em
universality} and the ability to entangle all components of the
system.

More precisely, we consider the following problem: what dynamics can
we produce with a specified two-body, $n$-qubit Hamiltonian, given the
ability to perform arbitrary local unitary operations on individual
qubits?  Under these conditions, we exhibit an explicit algorithm
which shows that {\em any} Hamiltonian which produces entanglement can
be used to efficiently simulate an arbitrary two-body dynamical
operation.  This holds even if the Hamiltonian {\em alone} is only
capable of producing a small amount of entanglement.

It follows that any entangling interaction, together with local
unitaries, is sufficient to perform universal quantum computation.
Our results thus confirm the folklore belief that it is the ability to
entangle that is the crucial element in quantum computation.

%
%

Substantial prior work has been done on universal operations, and many
specific sets of universal gates are
known\cite{Barenco95a,Nielsen00a}.  Our work differs from previous
work on the general requirements for universality in several
regards. Closest is the work in \cite{Deutsch95a} and \cite{Lloyd95a},
where it was shown that almost any two-qubit quantum gate is universal
for quantum computation.  This work focused on unitary gates rather
than continuous-time Hamiltonian evolution, and did not explicitly
determine which sets of unitary gates are universal.  Our work
explicitly determines which two-body Hamiltonians, together with the
additional requirement of local unitaries, are universal.
Furthermore, in~\cite{Deutsch95a} and~\cite{Lloyd95a} it was assumed
that gates could be independently applied to {\em any} pair of qubits
in the computer, and thus required the ability to turn on and turn off
interactions between different pairs of qubits.  By contrast, we
assume only a fixed entangling operation.

Our techniques make use of generalisations of standard nuclear
magnetic resonance (NMR) techniques for decoupling and
refocusing~\cite{Slichter90a,Ernst94a}.  Similar ideas have been
applied by~\cite{Jones99a,Leung00a} to the problem of efficiently
implementing coupled logic gates using a restricted class of
hamiltonians which arises naturally in NMR.

%
%

The structure of this Letter is as follows.  We begin with a precise
formulation of our goals and results.  A specific two-qubit example is
given to illustrate our techniques, and the general algorithm is
described for the case of an arbitrary two-qubit system. The
efficiency of the algorithm and the effect of errors are then
discussed.  We conclude by generalizing the algorithm to $n$-qubit
systems.

%
%

An arbitrary Hamiltonian on $n$ qubits can be given the {\em operator
expansion} \begin{eqnarray} \label{eq:op_exp} H =
\sum_{j_1,\ldots,j_n=0}^3 h_{j_1 \ldots j_n} \sigma_{j_1}
 \otimes \ldots \otimes \sigma_{j_n},
\end{eqnarray}
where the $h_{j_1 \ldots j_n}$ are real numbers and
$\sigma_1,\sigma_2,\sigma_3$ are the usual Pauli sigma matrices, with
$\sigma_0 \equiv I$ the identity.  Our discussion is restricted to the
case of time-independent Hamiltonians containing only one- and
two-body terms, that is, if $h_{j_1\ldots j_n} \neq 0$ then only one
or two of the $j_1,\ldots,j_n$ are not equal to zero.  If the
Hamiltonian contains a non-zero contribution to $\sigma_k \otimes
\sigma_l$ then we say the Hamiltonian {\em couples} systems $k$ and
$l$.  This focus on two-body Hamiltonians is a mild restriction as
most candidate systems for quantum information processing are of this
type.

%
%

Under what circumstances is it possible to produce entanglement
between an {\em arbitrary} pair of systems, even ones that are not
directly coupled by the Hamiltonian $H$?  Not surprisingly,
Hamiltonians which have terms coupling systems $k$ and $l$ can produce
entanglement between these systems.  We say that systems $k$ and $k'$
are {\em connected} if there is a sequence $(k,k_1,\ldots,k_m,k')$
such that each adjacent pair in the sequence is coupled by $H$.  It is
clear that if $k$ and $k'$ are not connected then no entanglement can
be created between them, and thus it is not possible to perform an
arbitrary unitary operation on the system.  Conversely, it follows
from our later discussion (and is {\em a priori} plausible) that if a
pair of systems is connected then it is possible to create
entanglement between them (cf. \cite{Cirac01a,Dur00d}).  This motivates our
definition of a {\em two-body entangling Hamiltonian} as a two-body
Hamiltonian such that all pairs of systems are connected.

%
%

The main result of this Letter is the following:
\begin{itemize}
\item[]
Let $H$ be a given two-body entangling Hamiltonian on $n$ qubits, and
let $K$ be a desired two-body Hamiltonian on $n$ qubits.  Then we have
an efficient algorithm to simulate evolution due to $K$ using only (a)
the ability to evolve according to $H$ and (b) the ability to perform
local unitaries on the individual qubits.
\end{itemize}
In particular, given such a Hamiltonian it follows that we can perform
an arbitrary two-qubit unitary gate on any specified pair of qubits.
Thus, by well-known universality results\cite{Barenco95a,Nielsen00a}, we
may efficiently perform any quantum computation.

%
%
Three elementary observations about Hamiltonian evolution
form the key to our methods:

\noindent (A) Imagine we can evolve according to the Hamiltonian $J$, and
perform unitary operations $U$ and $U^{\dagger}$.  Then it follows
from the identity $e^{-it UJU^{\dagger}}= U e^{-itJ}U^{\dagger}$
that we can exactly simulate evolution according to the Hamiltonian
$UJU^{\dagger}$.

\noindent (B) Imagine we can evolve according to Hamiltonians $J_1$ and
$J_2$.  Then we can simulate evolution due to $J_1+J_2$ for small
times $\Delta$, due to the approximate identity
\begin{eqnarray} \label{eq:B}
e^{-i\Delta (J_1+J_2)} \approx e^{-i \Delta J_1}e^{-i \Delta J_2}.
\end{eqnarray}
Initially we treat this identity as though it is exact, and analyze
the effect of errors later.

\noindent (C) Imagine we can evolve according to a Hamiltonian $J$.  Then,
by appropriate timing, we can exactly simulate evolution according to
$\lambda J$ for any $\lambda > 0$.  

%
%

The basic idea can be illustrated using a two-qubit example.  Suppose
we have the ability to evolve according to the two-qubit Hamiltonian:
\begin{eqnarray} \label{eq:example} H = Z \otimes I + 2 X \otimes Z +
Z \otimes Z, \end{eqnarray} where $X, Y$ and $Z$ are a convenient
shorthand for the Pauli sigma matrices.  $H$ couples the
two qubits, and is thus a two-body entangling Hamiltonian.  The first
step of our procedure is to show that $H$ and local unitaries can be
used to simulate evolution according to the largest coupling term in
$H$, in this case $X \otimes Z$.  This follows immediately
from the identity
\begin{eqnarray}
X \otimes Z = \frac{1}{4} (X \otimes I)H(X\otimes I)^{\dagger} + \frac{1}{4}H,
\end{eqnarray}
and our earlier observations~(A),~(B) and~(C).

%
%

Using the ability to simulate evolution by the Hamiltonian $X \otimes
Z$ we can easily obtain the ability to simulate a Hamiltonian which is
{\em any} product of Pauli matrices.  Products of the form $I \otimes
\sigma_j$ and $\sigma_j \otimes I$ follow immediately from our ability
to do local unitaries.  Products of the form $\sigma_j \otimes
\sigma_k$ follow from observation~(A) and the fact that $\sigma_j
\otimes \sigma_k = (U \otimes V) X \otimes Z (U \otimes V)^{\dagger}$
for appropriate single-qubit rotations $U$ and $V$.  It is easy to see
that observation~(A) also allows us to simulate terms of the form
$-\sigma_j \otimes \sigma_k$.  An arbitrary two-qubit Hamiltonian $K$
can be decomposed as a linear combination of products of Pauli
matrices, and thus by observations~(B) and~(C) may be simulated using
our ability to simulate $X \otimes Z$.  Thus $K$ may be simulated
using $H$ and local unitaries.

%
%
The general two-qubit case follows using similar techniques.  Suppose
$H$ is an entangling Hamiltonian and choose $r, s\neq 0$ such that
$|h_{r,s}|$ is maximized.  It is easy to verify that
\begin{eqnarray} \label{general_two_qubit}
\mbox{sgn}(h_{r,s})\sigma_r \otimes \sigma_s &&
	= \sum_{j\in\{0,r\},k \in \{0,s\}}
	\frac{(\sigma_j \otimes \sigma_k)H(\sigma_j \otimes 
	\sigma_k)^{\dagger}} {4 |h_{r,s}|}\nonumber
\\
- &&\,\frac{h_{r,0} \sigma_r \otimes I +h_{0,s} I \otimes \sigma_s
	+h_{0,0}I \otimes I}{|h_{r,s}|}.
\end{eqnarray}
Using observations (A), (B) and (C) it
follows that both $\sigma_r \otimes \sigma_s$ and $-\sigma_r \otimes
\sigma_s$ can be simulated using $H$ and local unitaries, and thus the
result follows for a general two-qubit Hamiltonian $K$.

%
%

In more detail, suppose we wish to simulate $K$ for a
non-infinitesimal time $t > 0$.  We have shown that we can approximate
evolution due to $K$ for a small time $\Delta$ by applying an
appropriate sequence of evolutions due to $H$ and local unitaries.
Such a simulation requires, in general, $36$ separate periods of
evolution due to $H$, interleaved by single-qubit unitary gates
applied to the two qubits.  To simulate the evolution due to $K$ over
a time $t$ we break the interval $t$ into $N$ increments of length
$\Delta \equiv t/N$, and perform the simulation of $K$ for each
increment, repeating the small time-step procedure $N$ times, for a
total of at most $36N$ separate periods of evolution due to $H$.

%
%

Let us turn to the sources of error inherent in our simulation
procedure.  The procedure uses only the observations (A), (B) and (C).
(A) and (C) are in principle exact, however the identity
Eq.~(\ref{eq:B}) used in (B) only holds approximately.  In order to
perform a good simulation of $K$ we therefore need to choose a
timestep $\Delta$ sufficiently small that Eq.~(\ref{eq:B}) is a good
approximation.

%
%

To do the error analysis, we introduce a measure quantifying how well
our simulated evolution approximates the desired evolution due to $K$.
That is, we wish to compare the unitary evolution $W'$ achieved by our
simulation with the unitary evolution $W = \exp(-iKt)$ that we wish to
simulate.  We use as our measure of error the operator norm of the
difference between $W$ and $W'$, $\| W-W'\|$, defined by $\|A\| \equiv
\max_{\psi: | \psi | = 1} | A |\psi \rangle |$.  This is physically
well-motivated since two operators $W$ and $W'$ are close according to
this norm if and only if the difference in their effects on an
arbitrary state is bounded by a small number.  The actual measure of
error used is not all that important, but we find it useful to demand
the following two properties, both of which are satisfied by the
operator norm: (1) stability under tensor product with ancilla
systems, that is, $\| A \| = \| I \otimes A \|$; and (2) invariance
under unitary transformations, that is, $\|A \| = \| V' A V \|$ for
any unitaries $V, V'$.  This latter property implies the {\em chaining
inequality} for any unitary operators $V_1,V_2,W_1,W_2$:
\begin{eqnarray}
\| V_1W_1-V_2W_2 \| \leq \|V_1-V_2\|+\|W_1-W_2\|.
\end{eqnarray}

%
%

We bound the errors induced by the approximation in Eq.~(\ref{eq:B})
using the inequality\cite{deRaedt96a}:
\begin{eqnarray}
\| e^{-i \tau (A_1+\ldots +A_m)}- e^{-i\tau A_1}\ldots e^{-i\tau A_m} \| 
\nonumber \\
\leq \frac{\tau^2}{2} \sum_{1 \leq j \leq k \leq m} \| [A_j,A_k] \|,
\end{eqnarray}
where $\tau$ is a positive real number and the $A_j$ are Hermitian
operators.  Applying this bound and the chaining property to the
procedure we've described gives
\begin{eqnarray} \label{eq:bound}
\| W'-W\| \leq C D^2t \Delta,
\end{eqnarray}
where $C$ is a constant which we can easily bound to be at most
$10^4$, and $D$ is a parameter determined by the properties of $H$ and
$K$ as follows.  Let $h \equiv \max_{i,j} |h_{i,j}|, k \equiv
\max_{i,j} |k_{i,j}|$, where $K$ has the operator expansion
$\sum_{i,j} k_{i,j} \sigma_i \otimes \sigma_j$.  Then $D \equiv
|hk/h_{r,s}|$.

%
%

The error bound Eq.~(\ref{eq:bound}) can be improved substantially in
several ways.  The linear dependence on $\Delta$ in
Eq.~(\ref{eq:bound}) is due to the technique used to simulate sums of
Hamiltonians, namely $e^{-i\Delta(J_1+J_2)} = e^{-i\Delta J_1} e^{-i
\Delta J_2} +O(\Delta^2)$.  Each simulation step thus contributes an
error $O(\Delta^2)$, and there are $t/\Delta$ such steps for a total
error $O(t\Delta)$.  Higher-order approximation techniques
\cite{deRaedt96a} can be used to obtain more accurate simulations. For
example, identities such as
\begin{eqnarray} \label{eq:higher-order}
e^{-i\Delta(J_1+J_2)} =
e^{-i\Delta J_1/2} e^{-i\Delta J_2} e^{-i\Delta J_1/2}+O(\Delta^3)
\end{eqnarray}
yield a cumulative error which is $O(t\Delta^2)$.  In general, an
approximation analogous to Eq.~(\ref{eq:B}) but accurate to order
$\Delta^k$ leads to a cumulative error $O(t\Delta^{k-1})$. The
tradeoff is such that higher-order approximations require the use of
somewhat more complicated gate sequences for each timestep.  In
practical applications, this additional complication must be balanced
against the improvement in accuracy to achieve optimal results.

%
%

A second way to improve the bound in Eq.~(\ref{eq:bound}) is to
leverage specific knowledge of the given and desired Hamiltonians.
For example, imagine that we have available the Hamiltonian of
Eq.~(\ref{eq:example}), and wish to simulate a controlled-{\sc not}
gate\cite{Nielsen00a}.  We can do this more efficiently than implied
by the identity in Eq.~(\ref{general_two_qubit}) by examining the
properties of the controlled-{\sc not}.  Up to an unimportant global
phase, the controlled-{\sc not} may be generated by applying the
Hamiltonian $I \otimes X + Z \otimes I - Z \otimes X$ for a time $t =
\pi/4$.  The terms in this Hamiltonian commute, so we have
controlled-{\sc not}~$= e^{-i (I \otimes X) t}e^{-i (Z \otimes I)t}
e^{i(Z\otimes X)t}$.  Thus, to simulate the controlled-{\sc not} for a
time $t$, it suffices to simulate evolution according to the
Hamiltonian $K = -Z \otimes X$, followed by local unitaries.  We
observe that \begin{eqnarray} K = (R\otimes RX) \frac{(X\otimes
I)H(X\otimes I)^{\dagger}+H}{4} (R\otimes RX)^{\dagger} \end{eqnarray}
where $R$ is the Hadamard gate\cite{Nielsen00a}, denoted here by $R$
instead of the usual $H$ to avoid confusion with the given
Hamiltonian.  Using the method outlined earlier gives a cumulative
error $8 t \Delta$.  If we wish to have an accuracy of $10^{-3}$ this
corresponds to roughly $10^4$ periods of evolution according to $H$,
interleaved with local unitaries.  This number of operations is
probably too large to be practical, however it is substantially better
than is obtained using the general bound Eq.~(\ref{eq:bound}).

%
%

Further improvement may be obtained by using the higher-order
approximation Eq.~(\ref{eq:higher-order}).  Using the operator norm,
simple algebra shows that the correction in
Eq.~(\ref{eq:higher-order}) may be bounded to order $\Delta^3$ by
$\frac{1}{6} \| J_1 \| \|J_2\| \left( \|J_1\|+2\|J_2\|
\right)\Delta^3$.  In this specific example, this reduces to
$\frac{1}{128} \|H\|^3 \Delta^3$ for a cumulative error of at most
$\frac{1}{128}\|H\|^3 t \Delta^2$.  Bounding $\|H\|$ by $\| Z \otimes
I \| + 2 \| X \otimes Z \| + \|Z \otimes Z\| = 4$ we see that the
cumulative error is at most $\frac{1}{2} t \Delta^2$.  Therefore, to
achieve an accuracy of $10^{-3}$ in our simulation of the
controlled-{\sc not} we need approximately $10^2$ periods of evolution
due to $H$, interleaved with local unitaries.  Further improvements
may be obtained by using better approximations than
Eq.~(\ref{eq:higher-order}).

%
%

The number of operations required to simulate an arbitrary unitary
operation can thus be substantial.  In practice, this disadvantage may
be offset by the advantages gained in using the natural coherent
interactions present in a system.  Furthermore, our results merely
provide a lower bound on the efficiency with which it is possible to
simulate an arbitrary unitary operation, and provide substantial
impetus to search for better methods in specific cases.

%
%

We now turn to the $n$-qubit case.  The basic idea is to reduce the
problem to the two-qubit case already solved.  We divide the system
into two parts, a {\em principal system $P$} consisting of two qubits
which are coupled by the Hamiltonian $H$, and the {\em remainder} of
the system, denoted $S$.  We use a technique generalizing the work in
\cite{Leung00a,Linden99d} that turns off all interactions between $P$
and $S$ and within $S$, leaving only the interactions present in $P$.
These interactions can then be used, as before, to simulate arbitrary
dynamics on the two qubits in $P$.  Thus it is possible to simulate
arbitrary dynamics on {\em any} two qubits coupled by the Hamiltonian
$H$.  Finally, an arbitrary interaction between qubits $k$ and $k'$
may be effected by performing a sequence of {\sc swap} gates between
the qubits connecting $k$ and $k'$, applying the desired interaction,
and then swapping back.

%
%

The first step is to decouple systems $P$ and $S$.  To do this, let
$X_S$ denote a tensor product of $X$ operators applied bitwise to all
the qubits in $S$.  Define $Y_S$ and $Z_S$ similarly.  Observe that
forming the Hamiltonian
\begin{eqnarray}
H' = \frac{1}{4}\left[
	H+X_S H X_S^{\dagger}+Y_S H Y_S^{\dagger}+Z_S H Z_S^{\dagger}
	\right]
\end{eqnarray}
leaves the Hamiltonian on $P$ invariant, but eliminates all coupling
terms between $P$ and $S$, and all single-system terms on $S$.

%
%

We now explain a recursive construction to eliminate all remaining
couplings in $S$.  First, we break the block $S$ into two blocks $S_0$
and $S_1$ of approximately equal size.  We decouple $S_0$ and $S_1$ by
forming the Hamiltonian
\begin{eqnarray}
H'' &&\, = \nonumber \\
&&\frac{1}{4}\left[
	H'+X_{S_0}H'X_{S_0}^{\dagger}+Y_{S_0}H'Y_{S_0}^{\dagger}+
	Z_{S_0}H'Z_{S_0}^{\dagger} \right].
\end{eqnarray}
Next, we break $S_0$ into two blocks $S_{00}$ and $S_{01}$ of
approximately equal size, and break $S_1$ into two blocks $S_{10}$ and
$S_{11}$ of approximately equal size.  We can decouple $S_{00}$ from
$S_{10}$, and $S_{01}$ from $S_{11}$ in a {\em single step} by forming
the Hamiltonian
\begin{eqnarray}
H'''&&\,=\frac{1}{4} \left[ H''
	+X_{S_{00}}X_{S_{10}} H'' X_{S_{00}}^{\dagger}
	X_{S_{10}}^{\dagger}\right. \nonumber \\
	+&& \left. 
	\vphantom{ H''
	+X_{S_{00}'}X_{S_{10}} H'' X_{S_{00}}^{\dagger}X_{S_{10}}^{\dagger}}
	Y_{S_{00}}Y_{S_{10}} H'' Y_{S_{00}}^{\dagger} Y_{S_{10}}^{\dagger}
	+ Z_{S_{00}}Z_{S_{10}}H'' Z_{S_{00}}^{\dagger}Z_{S_{10}}^{\dagger}
	\right].
\end{eqnarray}
We repeat this blocking procedure $\lceil \log_2(n-2) \rceil$ times to
decouple all the terms in $S$, leaving a sum over $O(4^{\log_2 n}) =
O(n^2)$ terms involving the conjugation of $H$ by local unitary
operations.

%
%

Thus, simulating a Hamiltonian $K$ applied to $P$ for a time $t$
requires the use of $O(n^2)$ periods of evolution due to $H$,
interleaved with local unitaries.  Using a similar error analysis to
that described earlier, and the stability property of the operator
norm, we find an error $O(n^2 t \Delta)$.  In practice it may be
possible to do substantially better by leveraging our knowledge of
specific systems, and using better approximations.


A number of problems will be addressed in future work, including: (a)
the extension of our results beyond the qubit model to
higher-dimensional systems; (b) the {\em
fault-tolerance}\cite{Preskill98b,Nielsen00a} of our simulation
techniques; (c) the optimization of our techniques for specific
systems; (d) the further study of the general requirements for
universal computation
(cf. \cite{DiVincenzo00a,Bacon01a,Knill01a,Raussendorf01a}).  For
example, it is likely interesting to impose restrictions on the class
of local unitary operations that may be applied during the
computation, perhaps adopting a {\em cellular automata} model in which
operations are applied nearly homogeneously across the entire system.

%
%

The results presented in this paper demonstrate that all two-body,
$n$-qubit entangling Hamiltonians are {\em equivalent} in the sense
that any such Hamiltonian can be used to efficiently simulate any
other with the aid of local unitary operations.  We conjecture that
the same result is {\em not} true for $k$-body Hamiltonians where
$k>2$.  In the case $k=3$, for example, it is possible to construct a
Hamiltonian which, alone, can only generate GHZ-type entanglement.  It
would be surprising if this Hamiltonian could be used to simulate a
Hamiltonian which produces W-type entanglement because these types of
entanglement appear to be fundamentally different\cite{Dur00b}.  It
would be of interest to determine, in general, what characteristics of
two sets of Hamiltonians determine whether they are equivalent.

We thank Carl~Caves, Andrew~Childs, Ike~Chuang, Chris~Dawson,
Ivan~Deutsch, and Tobias~Osborne for helpful discussions.

\end{multicols}

\end{document}